\documentclass[prl, superscriptaddress,showpacs,twocolumn,nofootinbib]{revtex4-1}

\usepackage[normalem]{ulem}
\usepackage{graphicx}
\usepackage{color}
\usepackage{verbatim}
\usepackage{subfigure}
\usepackage{amssymb}
\usepackage{amsmath}
\usepackage{comment}
\usepackage{chngcntr}
\usepackage{appendix}
\usepackage{hyperref}

\newcommand{\brak}[1]{\left\langle#1\right\rangle}

\usepackage{amsthm}
\usepackage{bm}

\theoremstyle{plain}

\begin{document}

\title{Time reversal symmetry breaking and odd viscosity in active fluids: \protect\\
Green-Kubo and NEMD results}
\author{Cory Hargus}
\email{hargus@berkeley.edu}
\affiliation{Department of Chemical and Biomolecular Engineering, University of California, Berkeley, CA, USA}

\author{Katherine Klymko}
% \email{KKlymko@lbl.gov}
\affiliation{Computational Research Division, Lawrence Berkeley National Laboratory, Berkeley, CA, USA}

\author{Jeffrey M. Epstein}
% \email{epstein@berkeley.edu}
\affiliation{Department of Physics, University of California, Berkeley, CA, USA}

\author{Kranthi K. Mandadapu}
\email{kranthi@berkeley.edu}
\affiliation{Department of Chemical and Biomolecular Engineering, University of California, Berkeley, CA, USA}
\affiliation{Chemical Sciences Division, Lawrence Berkeley National Laboratory, Berkeley, CA, USA}

\begin{abstract}
Active fluids, which are driven at the microscale by non-conservative forces, are known to exhibit novel transport phenomena due to the breaking of time reversal symmetry.
Recently, Epstein and Mandadapu~\cite{Epstein2019} obtained Green-Kubo relations for the full set of viscous coefficients governing isotropic chiral active fluids, including the so-called odd viscosity, invoking Onsager's regression hypothesis for the decay of fluctuations in active non-equilibrium steady states.
In this Communication, we test these Green-Kubo relations using molecular dynamics simulations of a canonical model system consisting of actively torqued dumbbells.
We find the resulting odd and shear viscosity values from the Green-Kubo relations to be in good agreement with values measured independently through non-equilibrium molecular dynamics (NEMD) flow simulations.
This provides a test of the Green-Kubo relations, and lends support to the application of the Onsager regression hypothesis in relation to viscous behaviors of active matter systems.

\noindent \textbf{Journal version:}  J. Chem. Phys. 152(20), 201102 (2020). \href{https://doi.org/10.1063/5.0006441}{10.1063/5.0006441}.
\end{abstract}

\maketitle

\noindent\textbf{\textit{Introduction.}}
Statistical physics has traditionally been concerned with systems at equilibrium. A natural generalization pursued by Onsager, Prigogine, de Groot and Mazur, and others is to consider systems that are globally out of equilibrium but that obey the local equilibrium hypothesis \cite{deGroot1951,Mazur1984,Onsager1931a,Onsager1931b,Prigogine1967}.
Such systems model transport phenomena allowing linear laws, such as those of Fourier and Fick, to be derived from the principles of equilibrium thermodynamics and statistical mechanics \cite{Onsager1931a,Onsager1931b,Kubo1957b}.
The physical origin of the non-equilibrium nature of these systems is driving at boundaries, as in a rod heated from one end or a channel connecting regions of different solute concentration.

A more radical departure from equilibrium is achieved in active matter systems, in which equilibrium is broken at the local level by non-conservative microscopic forces.
Such activity is known to modify existing phase behavior as well as give rise to qualitatively new dynamical phases, as in motility-induced phase separation~\cite{tailleur2008statistical,cates2015motility}. Similarly, activity not only modifies existing transport coefficients, but can lead to entirely new coefficients, such as the odd (or Hall) viscosity appearing in chiral active fluids~\cite{banerjee2017odd,ganeshan2017odd,souslov2019topological,Epstein2019, liao2019mechanism, Bradlyn2012, han2020statistical}.

\begin{figure*}[!t]
    \centering
    \includegraphics[width=.9\textwidth]{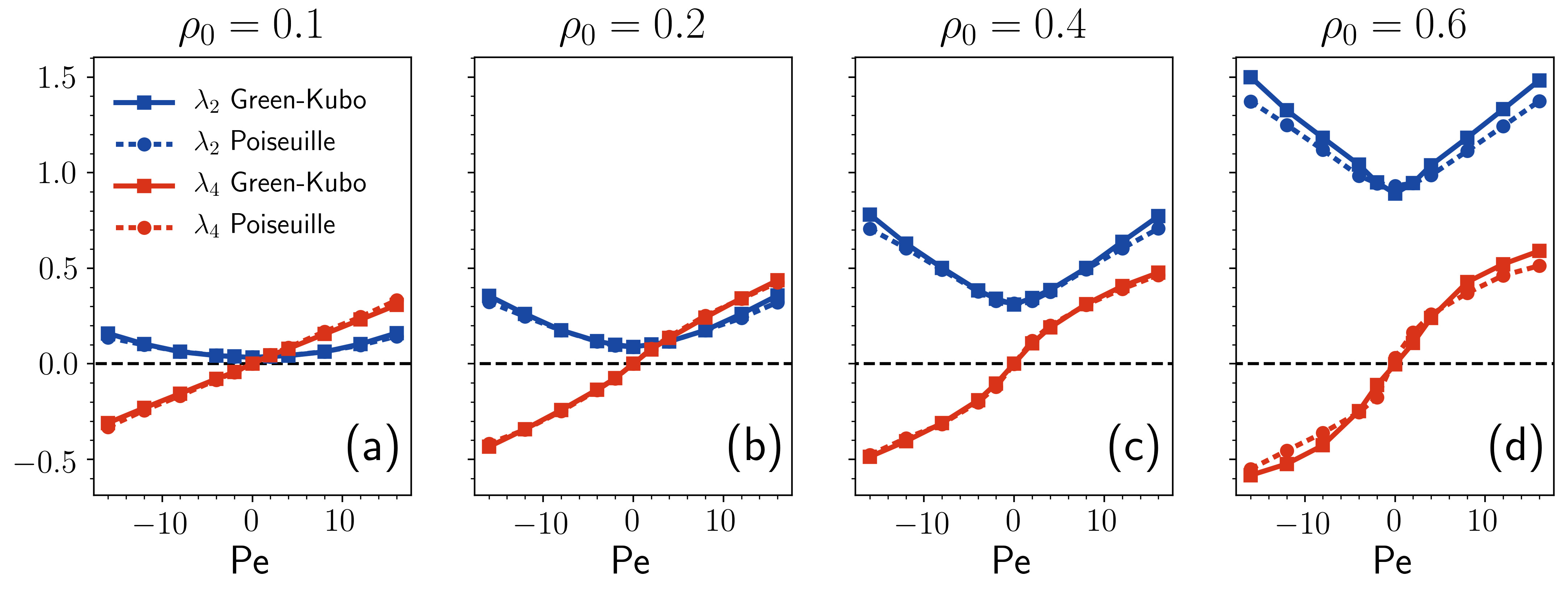}
    \caption{Comparison of shear viscosity ($\lambda_2$) and odd viscosity ($\lambda_4$) values obtained from the Green-Kubo relations (solid lines) with those obtained from periodic Poiseuille NEMD simulations (dashed lines). Error bars due to sampling convergence are smaller than the symbols. Figures~(a)-(d) show this comparison at densities $\rho_0 \in \{0.1, 0.2, 0.4, 0.6\}$, respectively. Each figure scans over $\mathrm{Pe} \in \{\text{-}16, \text{-}12, \text{-}8, \text{-}4, \text{-}2, 0, 2, 4, 8, 12, 16\}$.}
    \label{fig:gk_nemd_compare}
\end{figure*}

Recent work by Epstein and Mandadapu~\cite{Epstein2019} reveals that odd viscosity arises in two-dimensional chiral active fluids due to the breaking of time reversal symmetry at the level of stress correlations.
This is demonstrated by a set of Green-Kubo relations derived through the application of the Onsager regression hypothesis~\cite{Onsager1931a,Onsager1931b,Kubo1957b}.
In this Communication, we evaluate these Green-Kubo relations using molecular dynamics simulations of a model system composed of microscopically torqued dumbbells, finding them to be in good agreement with non-equilibrium molecular dynamics (NEMD) flow simulations across a wide range of densities and activities (Fig.~\ref{fig:gk_nemd_compare}).

\vspace{0.1in}

\noindent\textbf{\textit{Theory.}}
We begin by reviewing the continuum theory for two-dimensional viscous active fluids with internal spin.
This provides the setting for the derivation of Green-Kubo relations for viscosity coefficients in fluids breaking time reversal symmetry.
Because the chiral active dumbbell model considered in this paper is capable of storing angular momentum in the form of internal (\emph{i.e.}\ molecular) spin, we anticipate possible coupling between a velocity field $v_i$ and a spin field $m$.
These satisfy balance equations for linear and angular momentum, as proposed by Dahler and Scriven~\cite{dahler1961angular}:
\begin{align}
\label{eq:Cauchy_EOM}
\rho \dot{v}_i &= T_{ij,j} + \rho g_i \,, \\
\label{eq:Cauchy_EOM_spin}
\rho \dot{m} &= C_{i,i} - \epsilon_{ij} T_{ij} + \rho G\,.
\end{align}
$T_{ij}$ denotes the stress tensor and $C_i$ the spin flux, which accounts for transfer of internal angular momentum across surfaces.
The variables $g_i$ and $G$ denote body forces and body torques, respectively.
Finally note that the balance of angular momentum includes a term in which the two-dimensional Levi-Civita tensor $\epsilon_{ij}$ is contracted with the stress, so that the antisymmetric component of the stress may be nontrivial.
We use the notation $a_{,i}=\partial a/\partial x_i$.

The most general isotropic constitutive equations for viscous fluids relating $T_{ij}$ and $C_i$ to $v_i$, $m$ and their derivatives up to first order in two-dimensional systems are given by
\begin{align}
    \label{eq:general_constitutive_T}
    T_{ij} &= \eta_{ijkl} v_{k,l} + \gamma_{ij}m -p\delta_{ij} + p^*\epsilon_{ij}\,, \\
    \label{eq:general_constitutive_C}
    C_i &= \alpha_{ij}m_{,j}\,,
\end{align}
where $\eta_{ijkl}$, $\gamma_{ij}$ and $\alpha_{ij}$ are the viscous transport coefficients~\cite{Epstein2019}.
Here, $p$ and $p^*$ are hydrostatic contributions and are not constitutively related to $v_i$ and $m$.
The forms of equations~\eqref{eq:general_constitutive_T} and~\eqref{eq:general_constitutive_C} follow from a general representation theorem stating that any isotropic tensor can be expressed in a basis consisting of contractions of Kronecker tensors $\delta_{ij}$ and Levi-Civita tensors $\epsilon_{ij}$ and that, consequently, there exist no isotropic tensors of odd rank in two dimensions. Thus the transport coefficients may be expressed as
% Isotropy further allows the transport coefficients to be constrained to have the form
\begin{align}
    \label{eq:isotropic_constitutive_eta}
    \eta_{ijkl} &= \sum_{n =1}^6 \lambda_n s_{ijkl}^{(n)} \,, \\
    \label{eq:isotropic_constitutive_gamma}
    \gamma_{ij} &= \gamma_1\delta_{ij} + \gamma_2\epsilon_{ij} \,, \\
    \label{eq:isotropic_constitutive_alpha}
    \alpha_{ij} &= \alpha_1\delta_{ij} + \alpha_2\epsilon_{ij}\,,
\end{align}
where Table I contains the definitions of tensors $s^{(n)}_{ijkl}$.

\begin{table}[t!]
    \centering
    \setlength{\tabcolsep}{.5em}
    \renewcommand{\arraystretch}{1.2}
    \normalsize
    \begin{tabular}{|c|c|}\hline
    Basis Tensor    &   Components \\\hline
    $\mathbf{s}^{(1)}$& $\delta_{ij}\delta_{kl}$ \\
    $\mathbf{s}^{(2)}$& $\delta_{ik}\delta_{jl}-\epsilon_{ik}\epsilon_{jl}$ \\
    $\mathbf{s}^{(3)}$  &$\epsilon_{ij}\epsilon_{kl}$ \\
    $\mathbf{s}^{(4)}$& $\epsilon_{ik}\delta_{jl}+\epsilon_{jl}\delta_{ik}$ \\
    $\mathbf{s}^{(5)}$  &$\epsilon_{ik}\delta_{jl}-\epsilon_{jl}\delta_{ik}+\epsilon_{ij}\delta_{kl}+\epsilon_{kl}\delta_{ij}$ \\
    $\mathbf{s}^{(6)}$  &$\epsilon_{ik}\delta_{jl}-\epsilon_{jl}\delta_{ik}-\epsilon_{ij}\delta_{kl}-\epsilon_{kl}\delta_{ij}$ \\\hline
    \end{tabular}
    \caption{Basis of isotropic rank four tensors in two dimensions appearing in equation~\eqref{eq:isotropic_constitutive_eta}. Adapted from Ref.\ \cite{Epstein2019}.} \label{tab}
\end{table}

The coefficients $\gamma_n$ and $\alpha_n$ indicate the responses of the stress and spin flux tensors to spin and spin gradients.
$\lambda_1$ and $\lambda_2$ are the typical bulk and shear viscosities.
$\lambda_3$ is the rotational viscosity indicating resistance to vorticity and giving rise to an anti-symmetric stress, while $\lambda_4$ is the so-called odd viscosity quantifying response to shear with a tension or compression in the orthogonal direction.
$\lambda_5$ and $\lambda_6$ correspond to an anti-symmetric pressure from compression and isotropic pressure from vorticity, respectively.
Note that non-vanishing $\lambda_3$ or $\lambda_6$ violates objectivity (independence of stress from vorticity), while non-vanishing $\lambda_3$ or $\lambda_5$ violates symmetry of the stress tensor.

Using the conservation and constitutive equations~\eqref{eq:Cauchy_EOM}-\eqref{eq:Cauchy_EOM_spin} and \eqref{eq:isotropic_constitutive_eta}-\eqref{eq:isotropic_constitutive_alpha}, Ref.~\cite{Epstein2019} obtains a set of Green-Kubo relations for $\gamma_n$ and $\lambda_n$ via invocation of the Onsager regression hypothesis:
\begin{align}
\gamma_1&=\frac{1}{2\rho_0\nu}\delta_{ij}\epsilon_{kl} \mathcal{T}^{ijkl},\label{eq:GK1main}\\
\gamma_2&=\frac{1}{2\rho_0\nu}\epsilon_{ij}\epsilon_{kl}\mathcal{T}^{ijkl},\label{eq:GK2main}\\
\lambda_1+2\lambda_2+\lambda_3- \frac{\gamma_1\pi}{2\mu}+\frac{\gamma_2\tau}{2\mu}&=\frac{1}{2\rho_0\mu}\delta_{ik}\delta_{jl} \mathcal{T}^{ijkl},\label{eq:GK3main}\\
\lambda_4+\lambda_5+\lambda_6- \frac{\gamma_1\tau}{4\mu}-\frac{\gamma_2\pi}{4\mu}&=\frac{1}{4\rho_0\mu}\epsilon_{ik}\delta_{jl} \mathcal{T}^{ijkl},\label{eq:GK4main}\\
\lambda_5-\frac{\gamma_2 \pi}{4\mu}&=\frac{1}{8\rho_0\mu} \epsilon_{ij}\delta_{kl}\mathcal{T}^{ijkl},\label{eq:GK5main}\\
\lambda_3 +\frac{\gamma_2 \tau}{2\mu}&=\frac{1}{4\rho_0\mu}  \epsilon_{ij}\epsilon_{kl}\mathcal{T}^{ijkl}\label{eq:GK6main}\,.
\end{align}
$\mathcal{T}^{ijkl}$ is the integrated stress correlation function given by
\begin{equation}\label{eq:stress-correlator}
\mathcal{T}^{ijkl}=\int_{0}^\infty dt \langle \delta T_{ij}(t)\delta T_{kl}(0)\rangle.
\end{equation}
Note that the stress tensor in \eqref{eq:stress-correlator} is defined as a spatial average, as in the following section.
$\mu$, $\nu$, $\tau$, and $\pi$ are static correlation functions in the non-equilibrium steady state given by
\begin{align}
&\mu\delta_{ij}= \frac{1}{A^2}\int \brak{\delta v^i(\mathbf{x})\delta v^j(\mathbf{y})}d^2\mathbf{x}\,d^2\mathbf{y}\,,\label{eq:velocity-velocity}\\
&\pi= \frac{1}{A^2}\int (y^i-x^i)\brak{\delta v^i(\mathbf{x})\delta m(\mathbf{y})}d^2\mathbf{x}\,d^2\mathbf{y}\,,\\
&\tau= \frac{1}{A^2}\int \epsilon_{kr}(y^r-x^r)\brak{\delta m(\mathbf{x})\delta v^k(\mathbf{y})}d^2\mathbf{x}\,d^2\mathbf{y}\,,\\
&\nu= \frac{1}{A^2}\int \brak{\delta m(\mathbf{x})\delta m(\mathbf{y})}d^2\mathbf{x}\,d^2\mathbf{y}\,,\label{eq:nu}
\end{align}
respectively, where $A$ is the area of the system.
In particular, $\mu$ and $\nu$ can be regarded as measuring the effective translation and spin temperatures in the steady state.
For equilibrium systems, equipartition implies $\mu=\nu$ and $\pi=\tau=0$.
Lastly, the above Green-Kubo relations show that two of the transport coefficients, $\lambda_3$ and $\gamma_2$, are related by $2\lambda_3 = \gamma_2(\nu-\tau)/\mu$.

For the chiral active dumbbell fluid, the situation is further simplified.
As we will show in the following sections the absence of alignment interactions, \textit{i.e.} torque interactions acting at a distance between misaligned dumbbells, results in $\gamma_1 = \gamma_2 = 0$, effectively decoupling the velocity from the spin field and also setting $\lambda_3 = 0$.
Moreover, symmetry and objectivity of the stress tensor sets two more of the viscosity coefficients to zero, leaving
\begin{equation} \label{eq:constitutive}
\begin{split}
    \eta_{ijkl} = \lambda_1 \big(\delta_{ij} \delta_{kl}\big) + & \lambda_2 \big(\delta_{ik}\delta_{jl} - \epsilon_{ik}\epsilon_{jl} \big) \\
     & \hspace{0.2in} + \lambda_4 \big(\epsilon_{ik}\delta_{jl} + \epsilon_{jl} \delta_{ik} \big) \,.
    \end{split}
\end{equation}
These simplifications also allow us to write simplified Green-Kubo expressions for the shear viscosity
\begin{equation}
\begin{split}
    \label{eq:green_kubo_l2}
    \lambda_2 &= \frac{1}{4\rho_0 \mu} \int_0^\infty dt\ \langle (\delta T_{22}(t) - \delta T_{11}(t)) \\
    & \hspace{1.1in} (\delta T_{22}(0) - \delta T_{11}(0)) \rangle \,,
\end{split}
\end{equation}
and the odd viscosity
\begin{equation}\label{eq:green_kubo_l4}
\begin{split}
    \lambda_4 = \frac{1}{4 \rho_0 \mu} &\int_0^\infty dt \Big[ \langle \delta T_{11}(t) \delta T_{21}(0) \rangle - \langle \delta T_{11}(0) \delta T_{21}(t) \rangle  \\
    & \hspace{0.in}+ \langle  \delta T_{12}(t) \delta T_{22}(0) \rangle - \langle \delta T_{12}(0) \delta T_{22}(t) \rangle \bigg] \,,
\end{split}
\end{equation}
(see Appendix II for separating the coefficient $\lambda_2$ from \eqref{eq:GK3main}).
Equation~\eqref{eq:green_kubo_l4} shows that non-vanishing odd viscosity, \emph{i.e.}\ $\lambda_4\neq 0$, requires breaking time reversal symmetry at the level of stress correlation functions, thus breaking the Onsager reciprocal relations \cite{Onsager1931b, Epstein2019}.
Note that \eqref{eq:green_kubo_l2} is not the typical Green-Kubo expression used to calculate the shear viscosity.
However, it can also be rewritten for isotropic systems in the typical form, which are invariant under rotation as
\begin{equation}\label{eq:green_kubo_l2_rotated}
\lambda_2  = \frac{1}{\rho_0 \mu} \int_0^\infty dt\ \langle \delta T'_{12}(t) \delta T'_{12}(0) \rangle \,,
\end{equation}
using a transformation $\bm{T'} = \bm{R^T}\bm{T}\bm{R}$ corresponding to a rotation $\bm{R}$ of angle $\pi / 4$, for which $T_{12}' = \frac{1}{2}(T_{22} - T_{11})$.
The form in \eqref{eq:green_kubo_l2} is a result of the theory for the choice of the representation theorem for viscous transport coefficients using the basis $s^{(n)}_{ijkl}$.

In what follows, we evaluate the shear and odd viscosity Green-Kubo expressions at various densities and driving forces using molecular simulations of chiral active dumbbells in a non-equilibrium steady state.
We then subject the dumbbell system to non-uniform shearing flow, and evaluate the viscosity coefficients independently.
Such an analysis will provide support to both the application of Onsager's regression hypothesis to fluctuations in active non-equilibrium steady states and the ensuing Green-Kubo relations for viscous behaviors of active systems.

\begin{figure}[!htb]
    \centering
    \includegraphics[width=.42\textwidth]{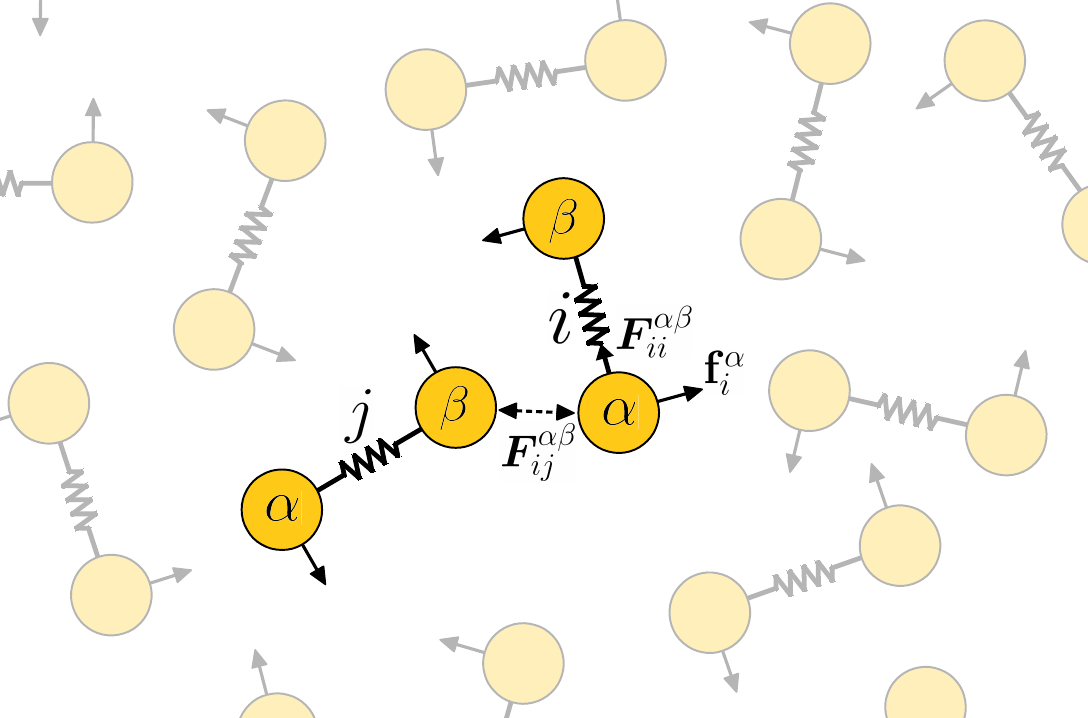}
    \caption{A two-dimensional fluid composed of chiral active dumbbells. In addition to interacting with its neighbors, each dumbbell is rotated counterclockwise by equal and opposite active forces $\mathbf{f}^\alpha_i$.}
    \label{fig:dumbbell_fluid}
\end{figure}
\vspace{0.1in}

\noindent\textbf{\textit{Microscopic model: Chiral active dumbbells.}} We consider a fluid composed of dumbbells subject to active torques \cite{klymko2017statistical}, as shown in Fig.~\ref{fig:dumbbell_fluid}.
Each dumbbell is composed of two particles of unit mass connected by a harmonic spring. The system evolves according to underdamped Langevin dynamics
\begin{align} \label{eq:Langevin_EOM}
    \begin{split}
    &\dot{\bm{x}}_i^\alpha = \bm{v}_i^\alpha \,, \\
    &\dot{\bm{v}}_i^\alpha = \sum_{j \beta}\bm{F}_{ij}^{\alpha\beta} + \mathbf{f}^\alpha_i +\bm{g}_i^\alpha - \zeta \bm{v}_i^\alpha + \bm{\eta}_i^\alpha\,,
    \end{split}
\end{align}
with indices $i,j \in [1,N]$ and $\alpha, \beta \in \{1,2\}$ running over dumbbells and particles, respectively.
Variables $\bm{x}_i^\alpha$ and $\bm{v}_i^\alpha$ represent atom positions and velocities.
$\zeta$ is the dissipative substrate friction and $T$ is the substrate temperature determining the variance of the random thermal force $\bm{\eta}_i^\alpha(t)$, modeled as Gaussian white noise affecting each particle independently such that, indicating vector components with indices $a$ and $b$, we have $\langle \eta_{ia}^\alpha(t) \eta_{jb}^\beta(t') \rangle = 2 k_\mathrm{B} T \zeta \delta(t-t') \delta_{ab} \delta_{ij} \delta_{\alpha \beta}$.
Particles in different dumbbells interact through a pairwise WCA potential \cite{Weeks71} (defined in Eq.~\eqref{eq:wca} of the Supplementary Material), resulting in interaction forces $\bm{F}_{ij}^{\alpha\beta}$.
The particles in a dumbbell are subjected to equal and opposite non-conservative active forces $\mathbf{f}^\alpha_i$, which satisfy $\mathbf{f}^1_i = -\mathbf{f}^2_i := \mathbf{f}_i$, and are always perpendicular to the bond vector $\bm{d}_i = \bm{x}_{i}^{1} - \bm{x}_{i}^{2}$.
This imposes an active torque at the level of individual dumbbells.
Finally, $\bm{g}_i^\alpha = \bm{g}(\bm{x}_i^\alpha)$ is an optional externally imposed body force, and will be employed later in Poisueille flow simulations to test the Green-Kubo relations.

Previous work~\cite{klymko2017statistical} used the Irving-Kirkwood procedure to coarse-grain the microscopic equations \eqref{eq:Langevin_EOM} and derive the equations of hydrodynamics, including balance of mass, linear momentum and angular momentum, as also employed in the context of measuring odd viscosity by~\cite{liao2019mechanism}.
This coarse-graining procedure yields expressions for the stress tensor in terms of molecular variables and active forces.
In particular, it is found that applying active forces at the microscale results in an asymmetric stress tensor at the continuum scale given by
\begin{align}\label{eq:irving_kirkwood}
    \bm{T} = \bm{T}^{\text{K}} + \bm{T}^{\text{V}} + \bm{T}^{\text{A}} \,,
\end{align}
where
\begin{align}\label{eq:irving_kirkwood-ind}
    \bm{T}^{\text{K}} &= -\frac{1}{A} \sum_{i ,\alpha} m_i^\alpha \bm{v}_i^\alpha \otimes \bm{v}_i^\alpha \,, \\
    \bm{T}^{\text{V}} & =  -\frac{1}{2A} \sum_{i,j,\alpha,\beta} \bm{F}_{ij}^{\alpha\beta} \otimes \bm{x}_{ij}^{\alpha\beta}\,, \label{eq:irving_kirkwood-ind-1} \\
    \bm{T}^{\text{A}} &= -\frac{1}{A} \sum_{i} \mathbf{f}_i \otimes \bm{d}_{i} \,, \label{eq:irving_kirkwood-ind-2}
\end{align}
denote the kinetic, virial, and active contributions, respectively.

The active force vector $\mathbf{f}_i$ is related to the unit bond vector $\hat{\bm{d}_i}$ by a rotation $\bm{R}$ of angle $\pi/2$, \emph{i.e.,}
\begin{equation}\label{eq:f_rot}
    \mathbf{f}_i = \mathrm{f} \bm{R} \hat{\bm{d}_i}
\end{equation}
For positive (negative) $\mathrm{f}$, the dumbbells rotate counter-clockwise (clockwise).
We find that the steady state time average of $\bm{T}^{\text{A}}$ is
\begin{align}
\begin{split}\label{eq:antisymmetric_stress}
\langle \bm{T}^{\text{A}} \rangle
& = -\rho_0 \langle \mathbf{f} \otimes \bm{d} \rangle \\
& = -\rho_0 \mathrm{f}d \langle \bm{R}\hat{\bm{d}} \otimes \hat{\bm{d}} \rangle
= \frac{\rho_0 \mathrm{f}d}{2}
    \begin{bmatrix}
   0  &
   1 \\
   \text{-}1 &
   0
   \end{bmatrix} \,,
\end{split}
\end{align}
where $d = \langle|\bm{d}|\rangle$ is the average bond length.
Because the dumbbells rotate with no preferred alignment, the antisymmetry of $\langle \bm{T}^{\text{A}} \rangle$ follows from replacing the time average with a uniformly weighted average over angles of rotation $\theta$.
For example,
\begin{equation}
     \langle \bm{R}\hat{\bm{d}} \otimes \hat{\bm{d}} \rangle_{21} = \langle \hat{d}_1 \hat{d}_1 \rangle = \frac{1}{2\pi} \int_0^{2\pi} d\theta\ \cos^2(\theta) = \frac{1}{2}
\end{equation}
while the diagonal elements are zero.
This shows that the antisymmetric hydrostatic-like term $p^*$ introduced in \eqref{eq:general_constitutive_T} arises in a non-equilibrium steady state of the active dumbbell model due to the presence of active rotational forces, and has the magnitude $p^* = \rho_0 \mathrm{f}d/2$.
We further relate $p^*$ to a non-dimensional P\'eclet number describing the ratio of active rotational forces to thermal fluctuations due to the substrate bath held at temperature $T$
\begin{equation}\label{eq:Peclet}
    \mathrm{Pe} = \frac{2\mathrm{f} d}{k_\mathrm{B}T} = \frac{4p^*}{\rho_0 k_\mathrm{B}T}.
\end{equation}
We use Pe as defined in \eqref{eq:Peclet} to vary the activity in the system when evaluating the transport coeffcients.

\vspace{0.1in}

\noindent\textbf{\textit{Green-Kubo calculations.}}
Steady-state molecular dynamics simulations~\cite{lammps} allow direct measurement of the integrated stress correlation functions $\mathcal{T}_{ijkl}$ defined in \eqref{eq:stress-correlator}, which are required for evaluation of the viscous transport coefficients using the Green-Kubo equations \eqref{eq:GK1main}-\eqref{eq:GK6main}~\footnote{Our simulation and analysis code is publicly available at https://github.com/mandadapu-group/active-matter.}.
We find that several of these coefficients vanish in the non-equilibrium steady states at all simulated activities and densities due to cancellations of the correlation functions (see Appendix Fig.~\ref{fig:all_correlators}).%(see Appendix Fig.~A.1).
In particular,
\begin{equation}
\epsilon_{ij}\epsilon_{kl}\mathcal{T}_{ijkl} = \delta_{ij}\epsilon_{kl}\mathcal{T}_{ijkl} = \epsilon_{ij}\delta_{kl}\mathcal{T}_{ijkl} = 0 \,.
\end{equation}
This immediately implies $\gamma_1 = \gamma_2 = \lambda_3 = \lambda_5 = \lambda_6 = 0$, so that the stress tensor is symmetric and objective. It now remains to evaluate the two non-trivial transport coefficients $\lambda_2$ and $\lambda_4$ using \eqref{eq:green_kubo_l2} and \eqref{eq:green_kubo_l4}.
For these coefficients, we compute the effective translation temperature as $(A\rho_0) \mu = m \langle (v_i^\alpha)^2 \rangle$, consistent with the stress tensor defined in \eqref{eq:irving_kirkwood}-\eqref{eq:irving_kirkwood-ind-2}.

\begin{figure}[!t]
    \centering
    \includegraphics[width=.48\textwidth]{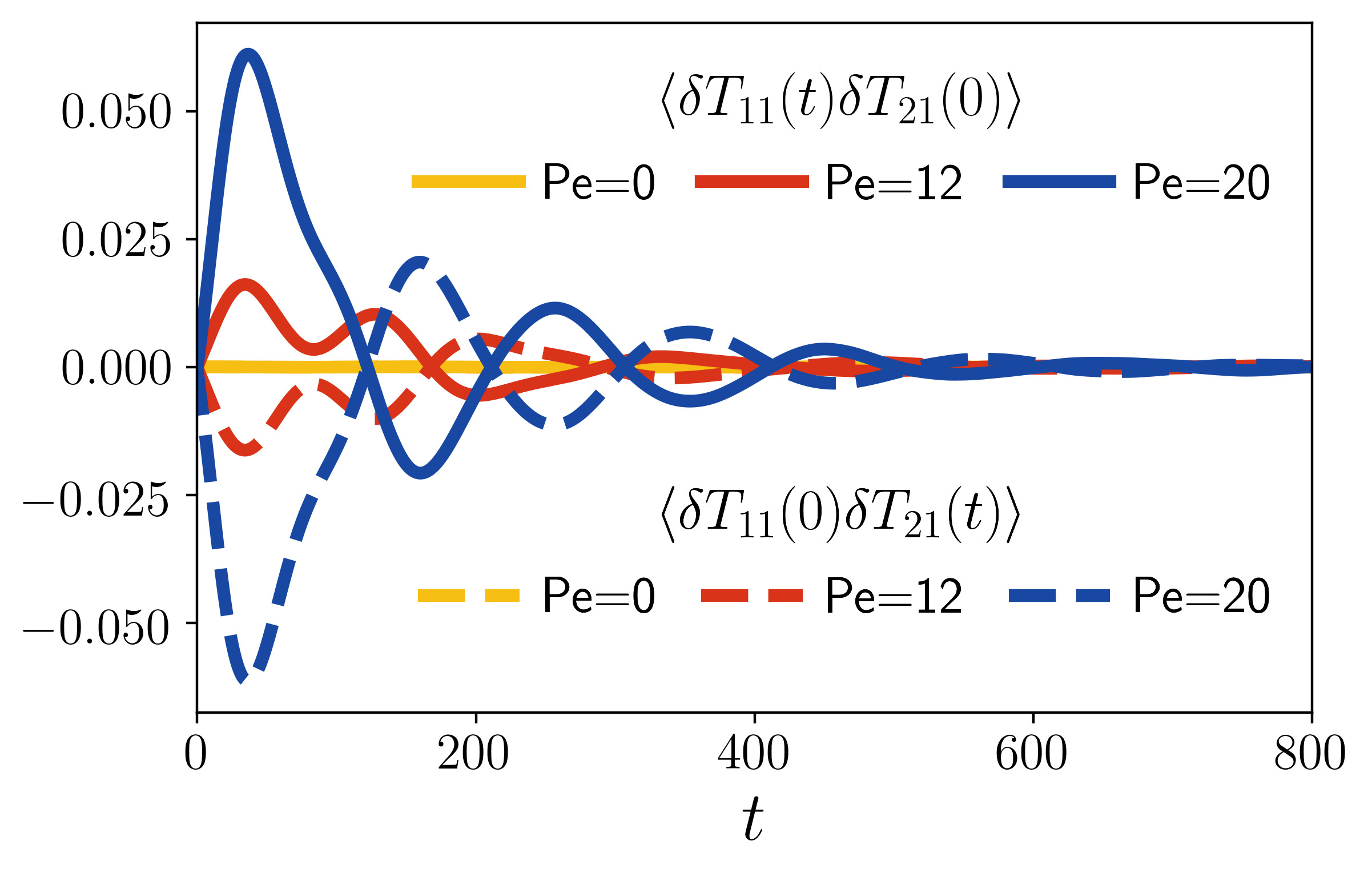}
    \caption{Stress correlation functions contributing to the odd viscosity ($\rho_0 = 0.4$). For Pe $\neq 0$ these correlation functions display time reversal antisymmetry, adding constructively to yield a nonzero odd viscosity.}
    \label{fig:odd_correlators}
\end{figure}

Figure~\ref{fig:odd_correlators} shows the stress correlation functions $\langle \delta T_{11}(t) \delta T_{21}(0) \rangle$ and $\langle \delta T_{11}(0) \delta T_{21}(t) \rangle$ for various $\mathrm{Pe}$.
These are typically zero for systems in equilibrium, but become nonzero in the chiral active dumbbell fluid for $\mathrm{Pe}\neq 0$.
In general, we find
\begin{equation}\label{eq:correlation-function-antisymmetry}
\begin{split}
    \langle \delta T_{11}(t) \delta T_{21}(0) \rangle &= - \langle \delta T_{11}(0) \delta T_{21}(t) \rangle  \\
    & = - \langle \delta T_{11}(\text{-}t) \delta T_{21}(0) \rangle \,,
    \end{split}
\end{equation}
where the final equality is due to stationarity.
The analogous equations are satisfied by $ \langle \delta T_{12}(t)  \delta T_{22}(0) \rangle$.
Due to this time reversal antisymmetry these correlation functions add constructively, yielding a non-vanishing odd viscosity from the Green-Kubo relation  \eqref{eq:green_kubo_l4}.

Figure~\ref{fig:gk_nemd_compare} shows the Green-Kubo estimates for $\lambda_2$ and $\lambda_4$ for various activities and for a range of low to high densities.
We find that the shear viscosity increases with density as well as with activity. The dependence of the odd viscosity on activity, while apparently linear at low density, becomes increasingly sigmoidal at high density.
Because the sign of Pe controls the direction of active rotation, the time reversal symmetry and antisymmetry, respectively, of $\lambda_2$ and $\lambda_4$ in equations~\eqref{eq:green_kubo_l2} and~\eqref{eq:green_kubo_l4} require that $\lambda_2$ must be an even function of Pe while $\lambda_4$ must be odd.
Note that the odd viscosity, as a non-dissipative transport coefficient, may be negative without introducing an inconsistency with the second law of thermodynamics.

\begin{figure}[!b]
    \centering
    \includegraphics[width=.4\textwidth]{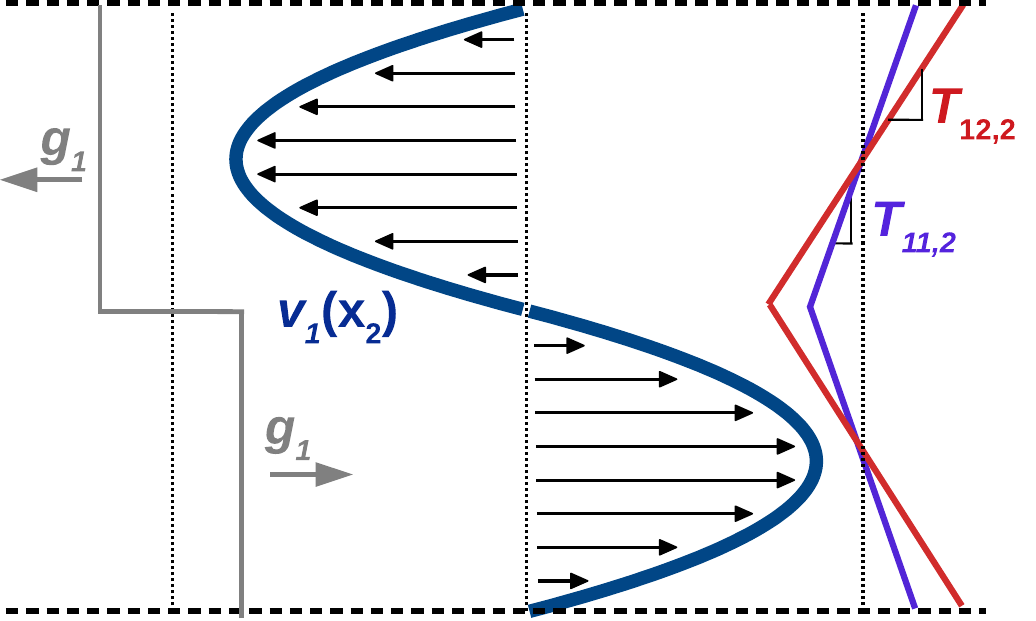}
    \caption{A schematic of the periodic Poiseuille non-equilibrium molecular dynamics (NEMD) simulation method.
    The top half of the system is subjected to a uniform body force to the left, and the bottom half to a uniform body force of equal magnitude to the right.
    This yields a parabolic velocity profile and, for odd viscous fluids, an atypical normal stress $T_{11}$.}
    \label{fig:pp}
\end{figure}

\vspace{0.1in}

\noindent\textbf{\textit{Poiseuille flow NEMD simulations.}}
To verify the values computed from the Green-Kubo formulas,  \eqref{eq:green_kubo_l2} and~\eqref{eq:green_kubo_l4}, we measure $\lambda_2$ and $\lambda_4$ independently via non-equilibrium molecular dynamics simulations.
To this end, we simulate plane Poiseuille-like flow via the inclusion of a nonzero body force $\bm{g}$ in \eqref{eq:Langevin_EOM} according to the periodic Poiseuille method~\cite{Backer2005}.
As depicted in Fig.~\ref{fig:pp}, we apply equal and opposite uniform body forces of magnitude $g_1$ in the $x_1$ direction across a rectangular channel of width $2L$, compatible with periodic boundary conditions.
In the following analysis, we consider only the bottom half of the system depicted in Fig.~\ref{fig:pp}, as the top half is symmetrically identical.

The setup in Fig.~\ref{fig:pp} represents a non-trivial boundary value problem, which not only yields non-uniform flows and non-uniform stresses, but also provides a stringent test for the expected constitutive behaviors of the active dumbbell fluid and the estimates of the transport coefficients obtained from Green-Kubo formulas.
The velocity profile and pressure profile for flow driven by a small, uniform body force can be solved analytically from the continuum theory, yielding
\begin{equation}
\label{eq:poiseuille_again-main}
v_1(x_2) = \frac{\rho_0 g_1}{2\lambda_2} x_2 (L-x_2)\,,
\end{equation}
and
\begin{equation}\label{eq:pressure-profile-main}
    p(x_2) = \frac{\lambda_4}{\lambda_2}\rho_0 x_2 g_1 + p_0 \,,
\end{equation}
respectively, where $p_0$ is an arbitrary reference pressure (see Appendix IV for the solution to the corresponding boundary value problem).
Our simulations of active dumbbell fluids are consistent with these profiles for various densities and activities (see Appendix Fig.~\ref{fig:poiseuille_profiles}).%(see Appendix Fig.~A.3).
Given the velocity and pressure profiles in \eqref{eq:poiseuille_again-main} and \eqref{eq:pressure-profile-main}, the shear and odd viscosities can be computed from the expressions
\begin{align}
    \label{eq:poiseuille_l2}
    \lambda_2 &= \frac{\rho_0 g_1 L^2}{12 \bar{v}} \,, \\
    \label{eq:poiseuille_l4}
    \lambda_4 &= \frac{T_{11,2}}{2 v_{1,22}} = - \frac{\lambda_2 T_{11,2}}{2 \rho_0 g_1}\,,
\end{align}
respectively, where $\bar{v} = \dfrac{1}{L}\int_0^L dx_2 \  v_1(x_2) $; see Appendix IV.  The slope of the stress component $T_{11}$ can be identified in molecular simulations using the Irving-Kirkwood expression \eqref{eq:irving_kirkwood}-\eqref{eq:irving_kirkwood-ind-2}.

The shear and odd viscosities calculated using this NEMD approach are found to be in agreement with the Green-Kubo predictions for a wide range of densities and P\'{e}clet numbers, see Fig.~\ref{fig:gk_nemd_compare}.

\vspace{0.1in}

{\noindent\textbf{\textit{Discussion.}}}
In this work, we have validated the non-equilibrium Green-Kubo formulas derived in Ref.\ \cite{Epstein2019}, using molecular dynamics simulations of the chiral active dumbbell model system to show that odd viscosity is a direct consequence of the breaking of time reversal symmetry at the level of stress fluctuations.
In doing so, we provide support for the application of the Onsager regression hypothesis to fluctuations about non-equilibrium steady states, which was used to derive these equations.
Complementary work by Han \textit{et al.}~\cite{han2020statistical} measures transport coefficients including the odd  viscosity in a different model system consisting of frictional granular particles, upon obtaining Green-Kubo relations identical in form to \cite{Epstein2019} using a projection operator formalism and finding similar agreement with NEMD measurements. Together with the present work, these results suggest broad applicability of these Green-Kubo relations in active fluids.
Future work entails understanding the microscopic origins of the functional dependence of the viscosities with density and activity.

\vspace{0.1in}

{\noindent\textbf{\textit{Supplementary Material}}}
See Appendix for details of the simulation methodology and derivations related to the Green-Kubo relations and Poiseuille-like flow in the presence of odd viscosity.

\vspace{0.1in}

{\noindent\textbf{\textit{Data availability}}}
Simulation data from this study are available from the corresponding author upon request.

\vspace{0.1in}

\noindent\textbf{\textit{Acknowledgements.}} C.H.
is supported by the National Science Foundation Graduate Research Fellowship Program under Grant No.\ DGE 1752814. K.K.M is supported by Director, Office of Science, Office of Basic Energy Sciences, of the U.S. Department of Energy under contract No. DEAC02-05CH11231.

\bibliography{ref}
\onecolumngrid

\newpage
\begin{center}
\textbf{Supplemental Information: Appendices I-IV}~\footnote{Note that all equations and figures appearing in the appendices are indexed with the prefix ``A''.   References without an ``A'' refer to the main text.} \\
\vspace{0.05in}
\end{center}

\def\thesection{\Roman{section}}
\def\thesubsection{\Roman{section}.\Roman{subsection}}
\def\thesubsubsection{\Roman{section}.\Roman{subsection}.\Roman{subsubsection}}
\renewcommand{\theequation}{A.\arabic{equation}}
\renewcommand{\thefigure}{A.\arabic{figure}}

%\appendix
%\counterwithin{figure}{subsection}
\setcounter{equation}{0}
\setcounter{footnote}{0}
\setcounter{section}{0}
\setcounter{figure}{0}
% Intro to appendix

\subsection{I. Simulation Details}
To investigate the viscous behavior of a fluid composed of self-spinning dumbbells, we perform molecular dynamics simulations in LAMMPS~\cite{lammps}, implementing our own modifications~\footnote{We have published our simulation and analysis code at https://github.com/mandadapu-group/active-matter.} to impose microscopic driving forces and compute the active stress $\bm{T}^{\text{A}}$.
All measured quantities in both the Green-Kubo and NEMD calculations are converged with respect to timestep and system size.

Particles interact with their non-bonded neighbors through a Weeks-Chandler-Andersen~\cite{Weeks71} potential defined by

\begin{equation}\label{eq:wca}
V_{ij}^\mathrm{WCA}(r) = \begin{cases}
      4\epsilon \bigg[ \big(\sigma/r\big)^{12} - \big(\sigma/r\big)^6 \bigg] + \epsilon & r < 2^{1/6}\sigma \\
      0 & r \geq 2^{1/6}\sigma \,. \\
   \end{cases}
\end{equation}
Here, $\sigma$, $\epsilon$ and particle mass $m$ are the characteristic length, energy, and mass scales, which are used to define the Lennard-Jones units system.
All numerical settings and results in this Communication are reported in Lennard-Jones units.
The two particles in a single dumbbell are held together by a harmonic potential $V(r) = \frac{1}{2}k(r-r_0)^2$ with spring constant $k=100$ and reference length $r_0 = 1$.

Dynamics are evolved according to underdamped Langevin dynamics \eqref{eq:Langevin_EOM} with bath temperature $T=1.0$ and friction $\zeta = 2.0$.
We apply the Langevin bath interactions only along the $x_2$ direction, so as not to impede flow in the $x_1$ direction, and employ these conditions in both Green-Kubo and periodic Poiseuille simulations.
We note that imposing bath interactions selectively along $x_2$ may lead to a violation of isotropy by aligning dumbbells along a preferred axis.
In all simulations, however, we check that dumbbells have no preferred alignment by measuring the departure of the bond angle of a dumbbell projected onto $[0, \pi/2]$ from the reference value of $\pi/4$:
\begin{equation}
    \delta \theta_i^+ = \arctan\bigg(\frac{|\mathbf{d}_i\cdot \mathbf{e}_2|}{|\mathbf{d}_i\cdot \mathbf{e}_1|}\bigg) - \frac{\pi}{4} \,.
\end{equation}
We find that in all simulations, $\max(|\langle \delta \theta_i^+ \rangle|) < 0.01 $ radians, where angle brackets indicate averaging in time and maximization is in space.
We also confirm that the density is indeed uniform in all periodic Poiseuille calculations. The relative spatial variation in the density is bounded in all simulations by $\big(\langle(\delta \rho)^2\rangle / \langle \rho^2 \rangle\big)^{1/2} < 0.1\% $.

\subsection{II. Green-Kubo Formula for Shear Viscosity}

\begin{figure}[!b]
    \centering
    \includegraphics[width=.9\textwidth]{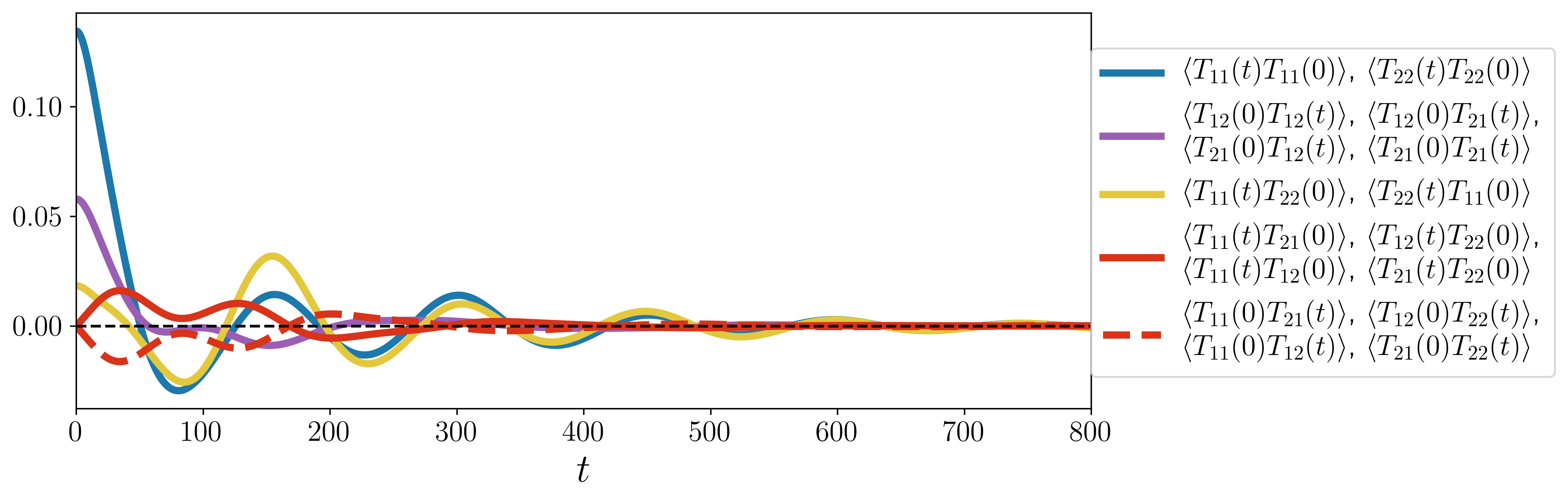}
    \caption{The sixteen stress correlation functions computed at $\rho_0 = 0.4$, $\mathrm{Pe} = 12$. Due to symmetries present in the chiral active dumbbell model, many of the correlation functions are identical, and are grouped as such. From this grouping, it is possible to ascertain that certain viscosity coefficients defined in \eqref{eq:GK1main}-\eqref{eq:GK6main} will vanish. For example, $\lambda_3$ depends on a sum of the correlation functions $\mathcal{T}_{1212} - \mathcal{T}_{1221} - \mathcal{T}_{2112} + \mathcal{T}_{2121}$. Here we see that these four correlation functions are identical, hence their sum will be zero. We further observe that the correlation functions contributing to the odd viscosity $\lambda_4$ go to zero in the static limit $t \rightarrow 0$, a consequence of the antisymmetry identified in \eqref{eq:correlation-function-antisymmetry}.}
    \label{fig:all_correlators}
\end{figure}

We also perform a derivation to obtain separate expressions for the shear and bulk viscosities. To this end, we begin with the following equation (also equation (127) in the SI of \cite{Epstein2019} in the absence of internal spin):
\begin{equation} \label{eq:LHS_and_RHS}
    k^j k^l \eta_{ijkl} = \frac{1}{\rho_0 \mu} k^j k^l \int_0^\infty dt\ \langle \delta T_{\mathbf{k}}^{ij}(t) \delta T_{\mathbf{-k}}^{kl}(0) \rangle
    = \frac{1}{\rho_0 \mu} k^j k^l \mathcal{T}^\mathbf{k}_{ijkl} \,,
\end{equation}
where
\begin{equation}
    \mathcal{T}^\mathbf{k}_{ijkl} = \int_0^\infty dt\ \langle \delta T_{\mathbf{k}}^{ij}(t) \delta T_{\mathbf{-k}}^{kl}(0) \rangle \,.
\end{equation}
Following \cite{Epstein2019}, we can obtain an equation for $\lambda_1$ and $\lambda_2$
\begin{equation}\label{eq:lambda1_and_2}
    \lambda_1 + 2\lambda_2 = \frac{1}{2\rho_0 \mu} \delta_{ik} \delta_{jk} \mathcal{T}^\mathbf{k}_{ijkl} \,,
\end{equation}
in the limit of $\mathbf{k} \rightarrow \mathbf{0}$.

To separate $\lambda_1$ from $\lambda_2$ we return to \eqref{eq:LHS_and_RHS} and contract both sides with $k^i k^k$ to obtain
\begin{equation}\label{eq:shear-visc-simplify-0}
    k^i k^j k^k k^l \eta_{ijkl} = \frac{1}{\rho_0 \mu} k^i k^j k^k k^l  \mathcal{T}^\mathbf{k}_{ijkl} \,.
\end{equation}
The resulting equation holds independently for any choice of $\mathbf{k}$ in the limit $\mathbf{k} \rightarrow 0$.
Now, we set $\mathbf{k} = k(\mathbf{e}_1 + \mathbf{e}_2)$ and $\mathbf{k} = k(\mathbf{e}_1 - \mathbf{e}_2)$ in \eqref{eq:shear-visc-simplify-0} and sum the resulting equations to obtain
\begin{equation}\label{eq:shear-visc-simplify-1}
    4\lambda_1 + 4\lambda_2 = \frac{1}{\rho_0\mu} \big(\mathcal{T}^\mathbf{k}_{1111} + \mathcal{T}^\mathbf{k}_{1122} + \mathcal{T}^\mathbf{k}_{1212} + \mathcal{T}^\mathbf{k}_{1221} + \mathcal{T}^\mathbf{k}_{2112} + \mathcal{T}^\mathbf{k}_{2121} + \mathcal{T}^\mathbf{k}_{2211} + \mathcal{T}^\mathbf{k}_{2222} \big)\,,
\end{equation}
which cannot be written in compact form as a contraction of Kronecker and Levi-Civita tensors with $\mathcal{T}^\mathbf{k}_{ijkl}$.
Subtracting \eqref{eq:shear-visc-simplify-1} from twice \eqref{eq:lambda1_and_2} and invoking the symmetry of the stress fluctuations gives
\begin{align}
\begin{split}
    \lambda_2 &= \frac{1}{4\rho_0 \mu}(\mathcal{T}^\mathbf{k}_{1111} - \mathcal{T}^\mathbf{k}_{1122} - \mathcal{T}^\mathbf{k}_{2211} + \mathcal{T}^\mathbf{k}_{2222}
                                            + \mathcal{T}^\mathbf{k}_{1212} - \mathcal{T}^\mathbf{k}_{1221} - \mathcal{T}^\mathbf{k}_{2112} + \mathcal{T}^\mathbf{k}_{2121}) \\
                     &= \frac{1}{4\rho_0 \mu}(\mathcal{T}^\mathbf{k}_{1111} - \mathcal{T}^\mathbf{k}_{1122} - \mathcal{T}^\mathbf{k}_{2211} + \mathcal{T}^\mathbf{k}_{2222})\,.
\end{split}
\end{align}
Finally, returning to the definition of $\mathcal{T}^\mathbf{k}_{ijkl}$ in \eqref{eq:LHS_and_RHS}, and taking the zero wavevector limit $\mathbf{k} \rightarrow 0$ yields
\begin{align}\label{eq:shear-visc-final}
\begin{split}
    \lambda_2 &= \frac{1}{4\rho_0 \mu} \int_0^\infty dt\ \langle (\delta T_{22}(t) - \delta T_{11}(t)) (\delta T_{22}(0) - \delta T_{11}(0)) \rangle \\
    &= \frac{1}{\rho_0 \mu} \int_0^\infty dt\ \langle \delta T_{12}(t) \delta T_{12}(0) \rangle \,,
\end{split}
\end{align}
where, in obtaining the last equality, we use material isotropy to make the stress transformation $\bm{T'} = \bm{R^T}\bm{T}\bm{R}$ corresponding to a two-dimensional rotation $\bm{R}$ of angle $\pi / 4$, for which $T_{12}' = \frac{1}{2}(T_{22} - T_{11})$.
The last equality in \eqref{eq:shear-visc-final} is the standard Green-Kubo relation for the shear viscosity. One may evaluate either of these expressions to compute the shear viscosity $\lambda_2$.

\subsection{III. Decomposed contributions to the viscosity coefficients from the Irving-Kirkwood stress tensor}
\begin{figure*}[!htb]
    \centering
    \includegraphics[width=.8\textwidth]{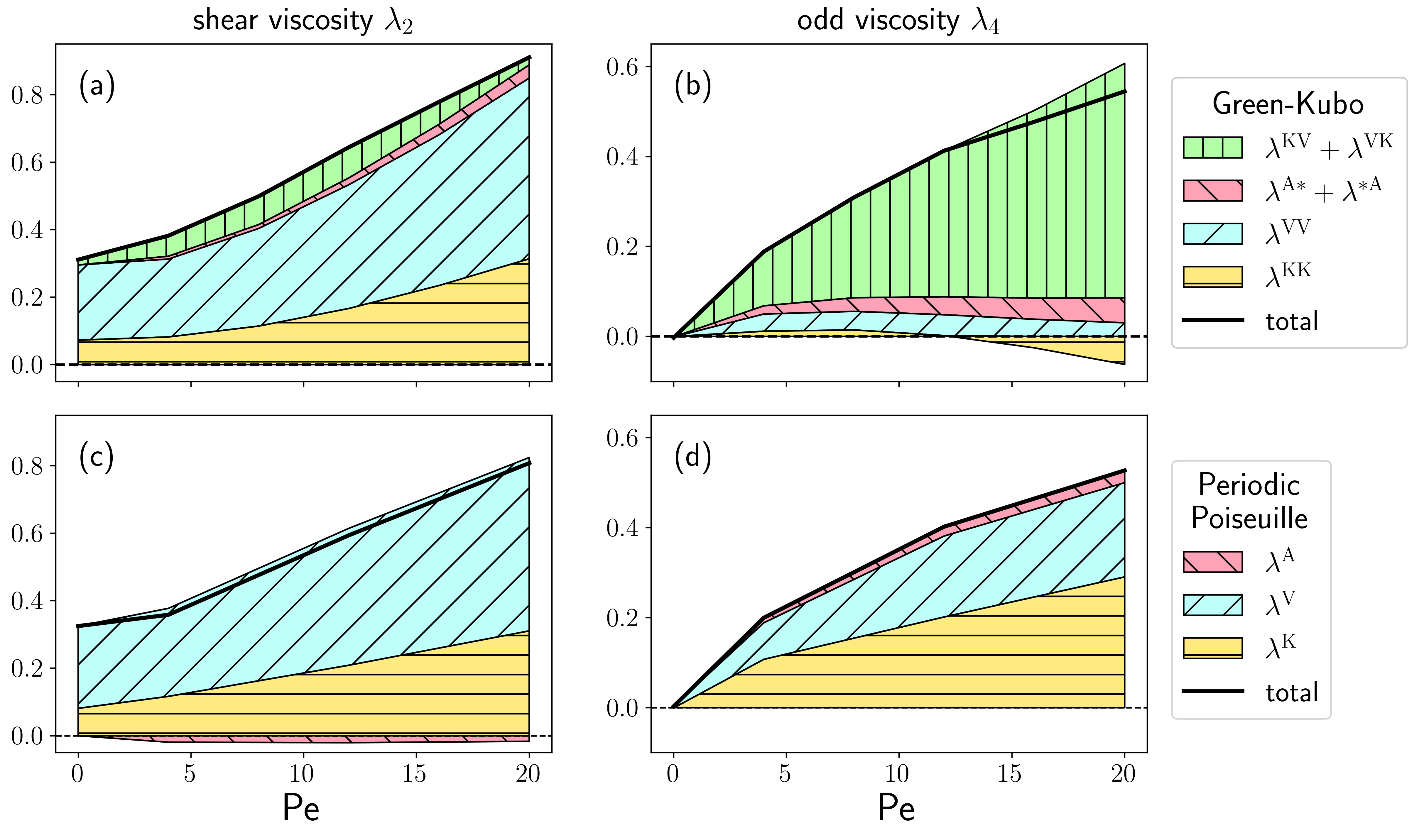}
    \caption{Components of the stress contributing to Green-Kubo and Poiseuille calculations of the shear and odd viscosity at $\rho_0 = 0.4$ as a function of Pe. Figures (a) and (b) are the component-wise contributions to $\lambda_2$ and $\lambda_4$, respectively, from Green-Kubo calculations according to the decompositions in \eqref{eq:decomp_gk_l2} and \eqref{eq:decomp_gk_l4}. Here, $\lambda^\mathrm{A*} + \lambda^\mathrm{*A} = \lambda^\mathrm{AK} + \lambda^\mathrm{AV} + \lambda^\mathrm{KA} + \lambda^\mathrm{VA} + \lambda^\mathrm{AA}$. Figures (c) and (d) are the component-wise contributions to the $\lambda_2$ and $\lambda_4$, respectively, in periodic Poiseuille calculations. The solid black line indicates the total viscosity coefficient, obtained by adding the shaded areas above $y=0$ and subtracting those below $y=0$.}
    \label{fig:stress_components}
\end{figure*}

The Irving-Kirkwood procedure provides a natural decomposition of the stress tensor into kinetic, virial, and active molecular contributions  \eqref{eq:irving_kirkwood}.
In Fig.~\ref{fig:stress_components}, we examine the component-wise stress contributions to the shear and odd viscosity in both Green-Kubo and periodic Poiseuille calculations.
The stress appears twice in the correlation functions entering the Green-Kubo equations \textit{via} \eqref{eq:stress-correlator}, thus there are nine components contributing to the Green-Kubo viscosity coefficients, which we label $\lambda^{\mathrm{KK}}$, $\lambda^{\mathrm{KV}}$, $\lambda^\mathrm{{KA}}$, $\lambda^\mathrm{{VK}}$, $\lambda^\mathrm{{VV}}$, $\lambda^\mathrm{{VA}}$, $\lambda^\mathrm{{AK}}$, $\lambda^{\mathrm{AV}}$ and $\lambda^{\mathrm{AA}}$.

From \eqref{eq:green_kubo_l2_rotated}, we define a decomposed shear viscosity as
\begin{equation}\label{eq:decomp_gk_l2}
    \lambda^{XY}_2 =  \frac{1}{\rho_0 \mu} \int_0^\infty dt\ \langle \delta T^X_{12}(t)\delta T^Y_{12}(0)\rangle \,,
\end{equation}
where $X, Y \in \{\mathrm{K},\mathrm{V},\mathrm{A}\}$ indicate the kinetic, virial and active parts.
Similarly, the odd viscosity from \eqref{eq:green_kubo_l4} may be decomposed as
\begin{equation}\label{eq:decomp_gk_l4}
    \lambda^{XY}_4 =  \frac{1}{4\rho_0 \mu} \int_0^\infty dt\ \langle \delta T^X_{ij}(t)\delta T^Y_{kl}(0)\rangle \epsilon_{ik} \delta_{jl} \,.
\end{equation}

For periodic Poiseuille calculations, the decompositions contributing to the viscous coefficients simply involve the choice of whether to use $\bm{T}^\mathrm{K}$, $\bm{T}^\mathrm{V}$, or $\bm{T}^\mathrm{A}$ in \eqref{eq:poiseuille_l2} and \eqref{eq:poiseuille_l4}, corresponding to $\lambda^\mathrm{K}$, $\lambda^\mathrm{V}$, and $\lambda^\mathrm{A}$, respectively. We observe that the active stress $\bm{T}^\mathrm{A}$ plays a small but not insignificant role in both $\lambda_2$ and $\lambda_4$ at $\mathrm{Pe} \neq 0$. Notably, the dominant Green-Kubo contributions to $\lambda_2$ are $\lambda^{\mathrm{KK}}$ and $\lambda^{\mathrm{VV}}$ while the cross correlations $\lambda^{\mathrm{KV}}$ and $\lambda^{\mathrm{VK}}$ are dominant in $\lambda_4$.

\subsection{IV. Periodic Poiseuille Simulation}

\begin{figure}[!b]
    \centering
    \includegraphics[width=.9\textwidth]{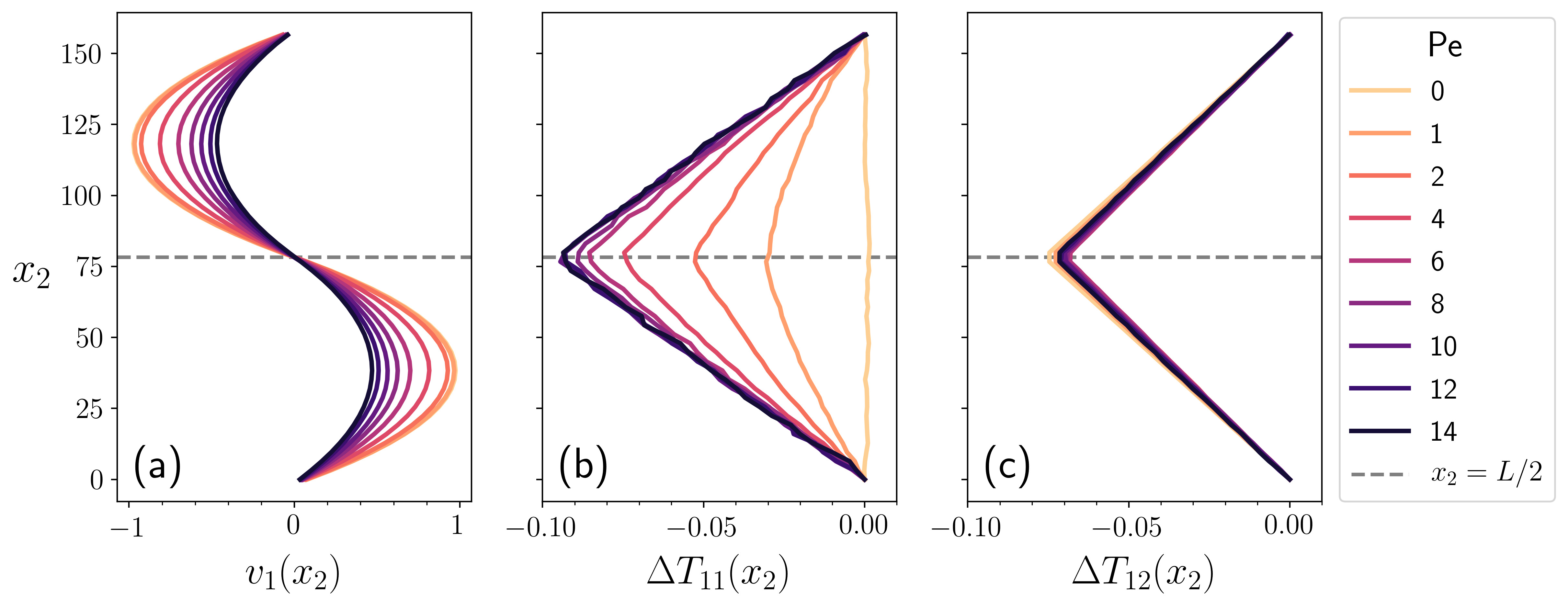}
    \caption{Time-averaged velocity and stress profiles from periodic Poiseuille simulations at $\rho_0 = 0.4$ over a range of Pe. Axes are chosen to be consistent with the schematic in Fig.~\ref{fig:pp}.
    Figure~(a) shows the velocity profile $v_1(x_2)$, where the increase in shear viscosity with increasing Pe is apparent, as described in \eqref{eq:si-lambda2}, in the decrease of the average velocity with increasing Pe.
    Figures~(b) and (c) show $\Delta T_{11}(x_2) = T_{11}(x_2) - \Delta T_{11}(0)$ and $\Delta T_{12}(x_2) = T_{12}(x_2) - \Delta T_{12}(0)$, respectively.
    Spatial variation in $T_{11}$ is seen to arise due to odd viscosity at $\mathrm{Pe} \neq 0$ as in \eqref{eq:si-lambda4}, while the slope of $T_{12}$ is unaffected by Pe, supporting the ansatz of constant $p^*$ used in \eqref{eq:poiseuille_again} and \eqref{eq:si-press}.
    }
    \label{fig:poiseuille_profiles}
\end{figure}

Non-equilibrium molecular dynamics simulations allow measurement of viscosity coefficients in direct analogy to experimental viscometry.
For the chiral active dumbbell fluid, $\gamma_1 = \gamma_2 = \lambda_3 = \lambda_5 = \lambda_6 = 0$, resulting in decoupling of the linear and angular momentum balances and leading to modified Navier-Stokes equations
\begin{equation}\label{eq:modified-NS-SI}
    \rho \dot{v}_i = \lambda_1 v_{k,ki} + \lambda_2 v_{i,jj} + \lambda_4 \epsilon_{ik} v_{k,jj} - p_{,i} + \epsilon_{ij}p^*_{,j} + \rho g_i \,,
\end{equation}
with bulk viscosity $\lambda_1$, shear viscosity $\lambda_2$, odd viscosity $\lambda_4$, pressure $p$, and body force $g_i$.

In the periodic Poiseuille simulations, we subject the system to equal and opposite body forces in the $x_1$ direction across a rectangular channel of width $2L$, as depicted in Fig.~\ref{fig:pp}.
In general, the non-uniform normal stress $\Delta T_{11}(x_2)$, due to the odd viscosity, may cause compression and extension of the fluid such that the steady state density is non-uniform in the $x_2$ direction.
Accordingly, we ensure that the body force $g_1$ driving the flow is sufficiently small in all simulations so that the density $\rho$ is well-approximated as constant, as described in Appendix I.
Therefore, we consider a steady state exhibiting incompressible flow, \emph{i.e.,}
\begin{equation}\label{eq:incompressible-flow}
    v_{i,i} = 0\,,
\end{equation}
and obtain the simplified constitutive and Navier-Stokes equations:
\begin{equation}\label{eq:simplified_constitutive}
    T_{ij} = \lambda_2 \big(v_{i,j} + v_{j,i} \big)
    + \lambda_4 \big(\epsilon_{ik} v_{k,j} + \epsilon_{jk} v_{i,k} \big)
    - p\delta_{ij} + p^*\epsilon_{ij} \,,
\end{equation}
and
\begin{equation}\label{eq:simplified_NS}
\rho_0 v_{i,j}v_j = \lambda_2 v_{i,jj} + \lambda_4 \epsilon_{ik} v_{k,jj} - p_{,i} + \epsilon_{ij}p^*_{,j} + \rho_0 g_i \,.
\end{equation}
where $\rho_0$ is the uniform reference density.

We now seek a steady state analytical solution for the velocity and pressure profiles of a fluid between two plates separated by a distance $L$, subjected to a body force $\mathbf{g} = (g_1,0)$, where $g_1$ is uniform in space.
The solution is analogous to that of a planar Poiseuille flow, with boundary conditions $v_i = 0$ at $x_2 = 0$ and $x_2 = L$.
Using the ansatz $v_1 = v_1(x_2)$, $v_2 = 0$, $p = p(x_2)$, and $p^* = \text{const}$, conditions which are observed in all non-equilibrium simulations considered in this study, one may find the steady state solution to be
\begin{equation}
\label{eq:poiseuille_again}
v_1(x_2) = \frac{\rho_0 g_1}{2\lambda_2} x_2 (L-x_2)\,,
\end{equation}
and
\begin{equation}\label{eq:si-press}
    p(x_2) = \frac{\lambda_4}{\lambda_2}\rho_0g_1 x_2 + p_0 \,,
\end{equation}
where $p_0$ is an arbitrary reference pressure.

We see that the steady state velocity profile is identical to the usual solution for planar Poiseuille flow, remaining unaffected by odd viscosity.
In fact it is always true that odd viscosity does not appear in the velocity profile in incompressible flows with no-slip boundary conditions~\cite{ganeshan2017odd}.
The odd viscosity does appear, however, in a pressure gradient arising in the $x_2$-direction to maintain the no-penetration condition at the walls, \textit{i.e.} to prevent flow in the $x_2$-direction.
Our active dumbbell fluid simulations show parabolic velocity profiles consistent with \eqref{eq:poiseuille_again} and \eqref{eq:si-press} when subjected to equal and opposite body forces as shown in Fig.~\ref{fig:pp}.

Integrating the velocity profile to get an average velocity $\bar{v} = \dfrac{1}{L}\int_0^L\  v_1(x_2) dx_2$, we obtain a convenient expression for computing the shear viscosity $\lambda_2$ in molecular simulations:
\begin{equation}\label{eq:si-lambda2}
    \lambda_2 = \frac{\rho_0 g_1 L^2}{12 \bar{v}} \,.
\end{equation}
As noted above, $\lambda_4$ does not appear in the velocity but in the stress \eqref{eq:simplified_constitutive}. For the velocity profile \eqref{eq:poiseuille_again},
\begin{equation}
    T_{11} = -p + \lambda_4 v_{1,2} \,,
\end{equation}
which results in
\begin{equation}\label{eq:T11,2}
    T_{11,2} =  -p_{,2} + \lambda_4 v_{1,22}\,.
\end{equation}
Using \eqref{eq:simplified_NS} in the $x_2$-direction, one may reduce \eqref{eq:T11,2} to
\begin{equation}\label{eq:T11,2-1}
    T_{11,2} =  2\lambda_4 v_{1,22} = -2\lambda_4\frac{\rho_0 g_1}{\lambda_2}\,.
\end{equation}
Finally, rearranging \eqref{eq:T11,2-1}, $\lambda_4$ is obtained in terms of the slope of $T_{11}$ as
\begin{equation}\label{eq:si-lambda4}
    \lambda_4 = \frac{T_{11,2}}{2 v_{1,22}} = - \frac{\lambda_2 T_{11,2}}{2 \rho_0 g_1}\,.
\end{equation}
where $T_{11}$ can be calculated using the Irving-Kirkwood formula \eqref{eq:irving_kirkwood} for the active dumbbell fluid.

\end{document}